\documentclass[10pt,twocolumn,letterpaper]{article}

\usepackage[pagenumbers]{cvpr} 

\usepackage[dvipsnames]{xcolor}

\usepackage[utf8]{inputenc} 
\usepackage[T1]{fontenc}    
\usepackage{url}            
\usepackage{booktabs}       
\usepackage{amsfonts}       
\usepackage{nicefrac}       
\usepackage{microtype}      
\usepackage[dvipsnames]{xcolor}         
\usepackage{multirow}
\usepackage{multicol}
\usepackage{pgfplots}
\usepackage{svg}
\usepackage{tikz}

\usepackage{overpic}
\usetikzlibrary{calc,patterns,angles,quotes}
\usepackage{pgfplots,gincltex}
\pgfplotsset{compat=1.18}
\usepgfplotslibrary{colormaps}
\usepackage{tikz-3dplot}
\pgfplotsset{compat=newest}
\usepackage{amsmath}

\definecolor{cvprblue}{rgb}{0.21,0.49,0.74}
\definecolor{lightboldcolor}{gray}{0.6}
\usepackage[pagebackref,breaklinks,colorlinks,allcolors=cvprblue]{hyperref}

\newcommand{\lightbold}[1]{\textbf{\textcolor{lightboldcolor}{#1}}}

\def\paperID{212} 
\def\confName{3DV\xspace}
\def\confYear{2026\xspace}

\title{Laplace-Beltrami Operator for Gaussian Splatting}

\author{  $\text{Hongyu Zhou}^{1,2}$ \hspace{16pt}  $\text{Zorah Lähner}^{1,2}$ \\
  $^{1}\text{University of Bonn}$ \hspace{16pt} $^{2}\text{Lamarr Institute for Machine Learning and Artificial Intelligence}$\\
   $\texttt{\{hzhou, laehner\}@uni-bonn.de}$ \hspace{16pt} $\texttt{lamarr-institute.org}$
}

\begin{document}
\maketitle
\begin{abstract}
With the rising popularity of 3D Gaussian splatting and the expanse of applications from rendering to 3D reconstruction, the need for geometry processing methods tailored directly to this representation becomes increasingly apparent. 
While existing approaches convert the centers of Gaussians to a point cloud or mesh to use them in existing algorithms, this conversion might discard valuable information present in the Gaussian parameters or introduce unnecessary computational overhead. 
Additionally, Gaussian splatting tends to contain a large number of outliers that, while not affecting the rendering quality, need to be handled correctly to not produce noisy results in geometry processing applications. 
In this work, we present a novel framework that operates directly on Gaussian splatting representations for geometry processing tasks. 
Our work introduces a graph-based outlier removal designed for Gaussian distributions as well as a formulation to compute the Laplace-Beltrami operator, a widely used tool in geometry processing, directly on Gaussian splatting.
Both use the Mahalanobis distance to account for the anisotropic nature of Gaussians. 
Our experiments show superior performance to the point cloud Laplacian operator and competitive performance to the traditional Laplacian operator computed on a mesh, while avoiding the need for intermediate representation conversion.

{\bf Website}: \href{https://zero-4869.github.io/LBO4GS/}{zero-4869.github.io/LBO4GS}
\end{abstract}
\section{Introduction}
\label{sec:intro}

3D Gaussian Splatting (3DGS)~\cite{kerbl3Dgaussians} has recently revolutionized 3D scene representations. By representing complex scenes as a set of 3D Gaussians, it achieves photorealistic results for novel view synthesis while allowing much more efficient training and real-time rendering compared to NeRF-based approaches~\cite{barron2022mip}. 
Even though it was originally proposed for rendering applications, it is now also popular as a general-purpose 3D representation~\cite{Huang2DGS2024,GaoMeshGaussian2024}. 
Despite its success in rendering, most 3DGS methods face fundamental challenges when applied to geometry processing tasks.
3DGS optimization is usually about visual fidelity but fails to produce geometrically accurate surfaces,
leading to recent work focusing on surface reconstruction instead~\cite{Huang2DGS2024,chen2023neusgneuralimplicitsurface}. 
There are two main strategies in this direction: 
methods that pull the 3D Gaussians to the suspected surface and then reconstruct a mesh via Poisson reconstruction~\cite{guedon2023sugar,kazhdan2006poisson} or TSDF~\cite{chen2024pgsr} for regularization, and
dimensionality-constraint approaches that restrict the Gaussians to be 2-dimensional which captures the 2-manifold structure of 3D objects~\cite{Huang2DGS2024} . 

As the geometric reconstruction accuracy improves, the need for traditional geometry processing tools that operate natively on Gaussian splatting arises.
Recent work has demonstrated the potential direct manipulation through deformation energies~\cite{huang2024sc} and semantic segmentation~\cite{cen2023saga} on Gaussian splats.
In this paper, we will expand this direction with a traditional operator from geometry processing, particularly the Laplace-Beltrami operator (LBO), which is often called the "Swiss Army knife" of geometry processing.
The LBO serves as an important part in applications ranging from mesh smoothing and deformation to spectral analysis and shape matching~\cite{sorkine2005laplacian,ovsjanikov2012functional,weber2024finsler}. 

One core information in the LBO is the connectivity information between points, which is by design missing in Gaussian splatting. 
A possible solution is to convert the 3DGS representation into a mesh, but this increases vertex count and computational overhead while often producing both over-smoothed details and noisy artifacts.
Alternatively, treating Gaussians as points clouds by only using their centers discards valuable geometric information encoded in the covariance matrices and suffers from poor performance due to the high density of outliers.
The outliers stem from the image-based optimization that naturally does not depict the inside of objects, and low-opacity Gaussians that contribute minimally to the rendering quality but add noise to the point cloud.
As a consequence, point cloud methods fail to capture the geometric structure implicitly encoded in the Gaussian splatting representation. 

\begin{figure*}[htb]
    \centering
    \setlength\tabcolsep{1pt}
    \begin{tabular}{cccc}
        \footnotesize{Reconstructed mesh} & \footnotesize{Point cloud Laplacian} & \footnotesize{Ours (Euclidean)} & \footnotesize{Ours (Mahalanobis + Normal)}  \\
         \includegraphics[width=0.24\linewidth]{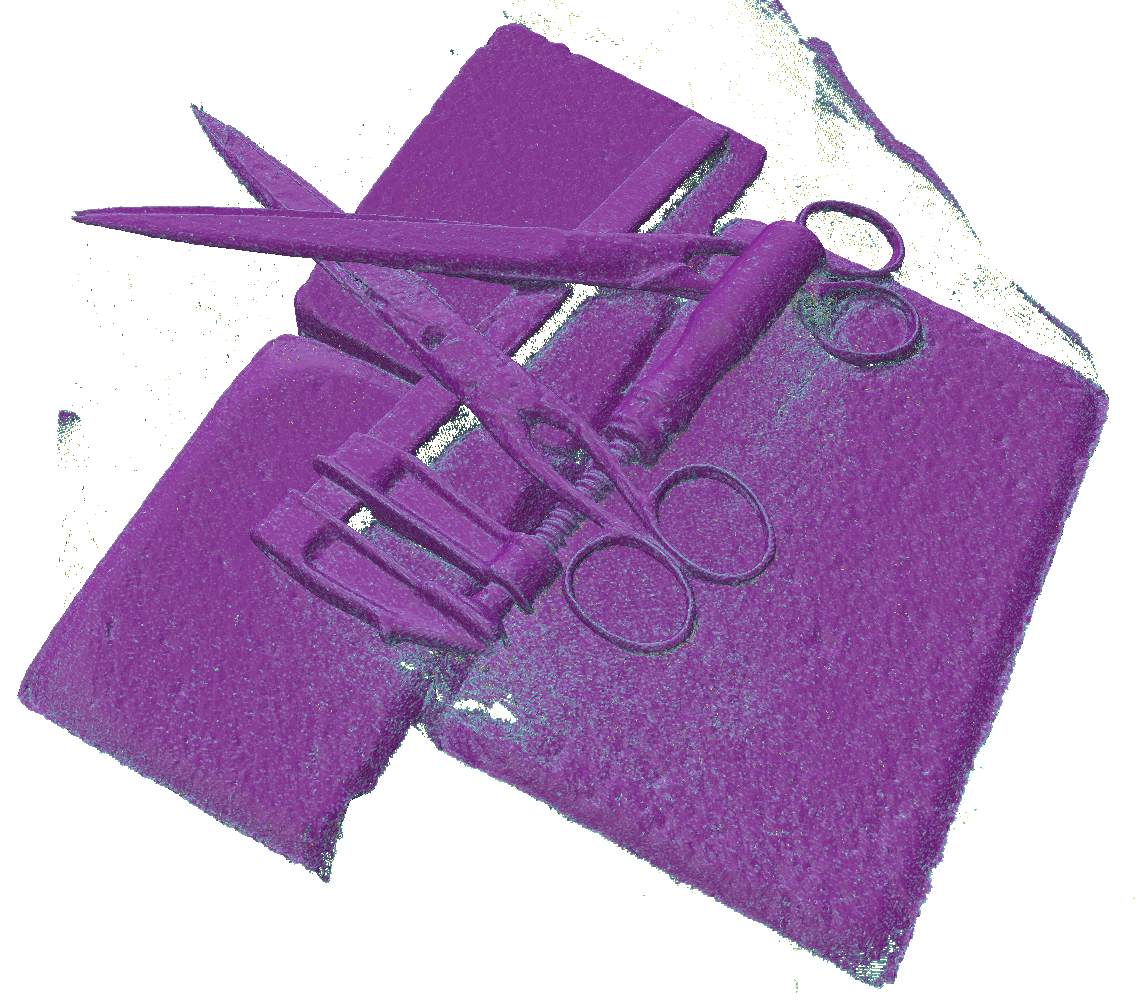} & \includegraphics[width=0.24\linewidth]{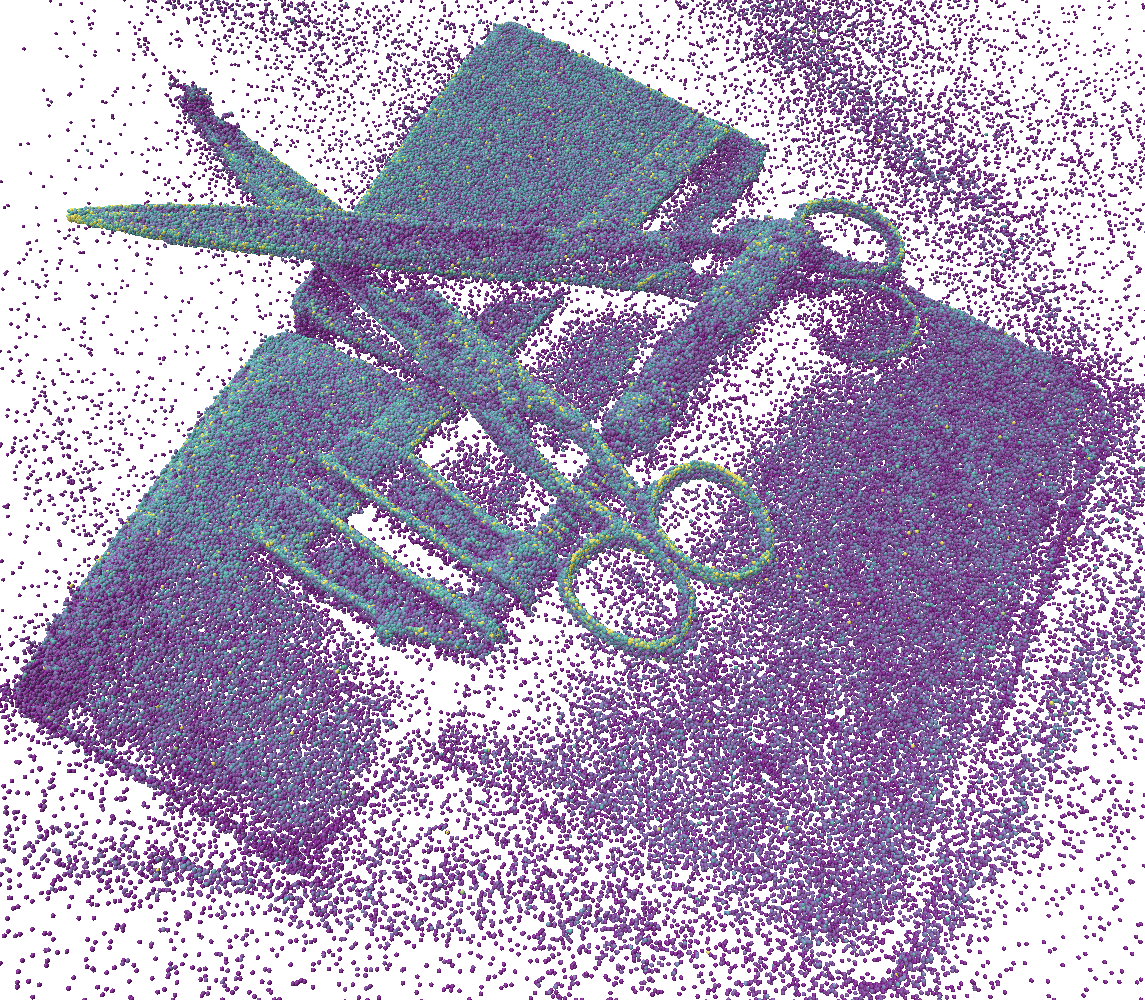} & \includegraphics[width=0.24\linewidth]{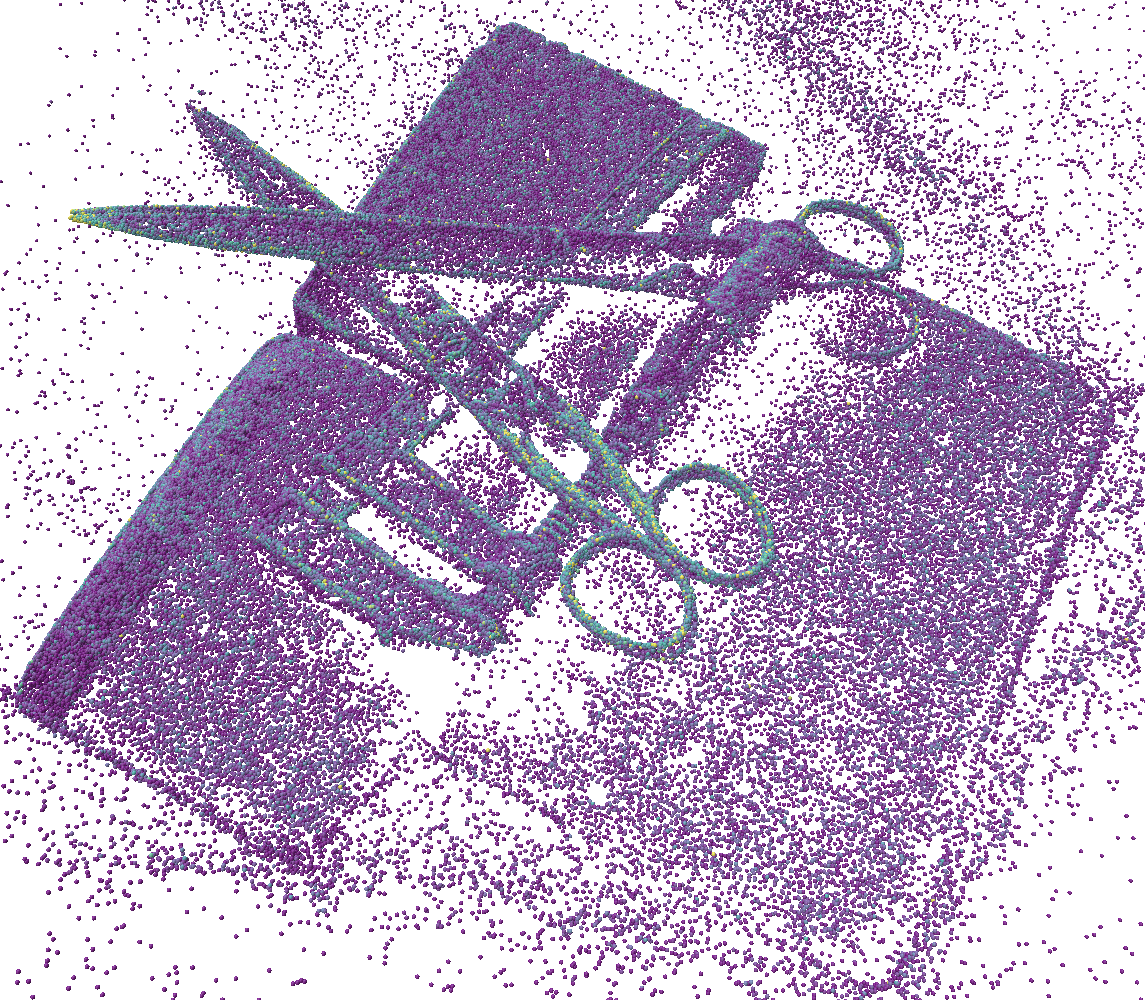} & \includegraphics[width=0.24\linewidth]{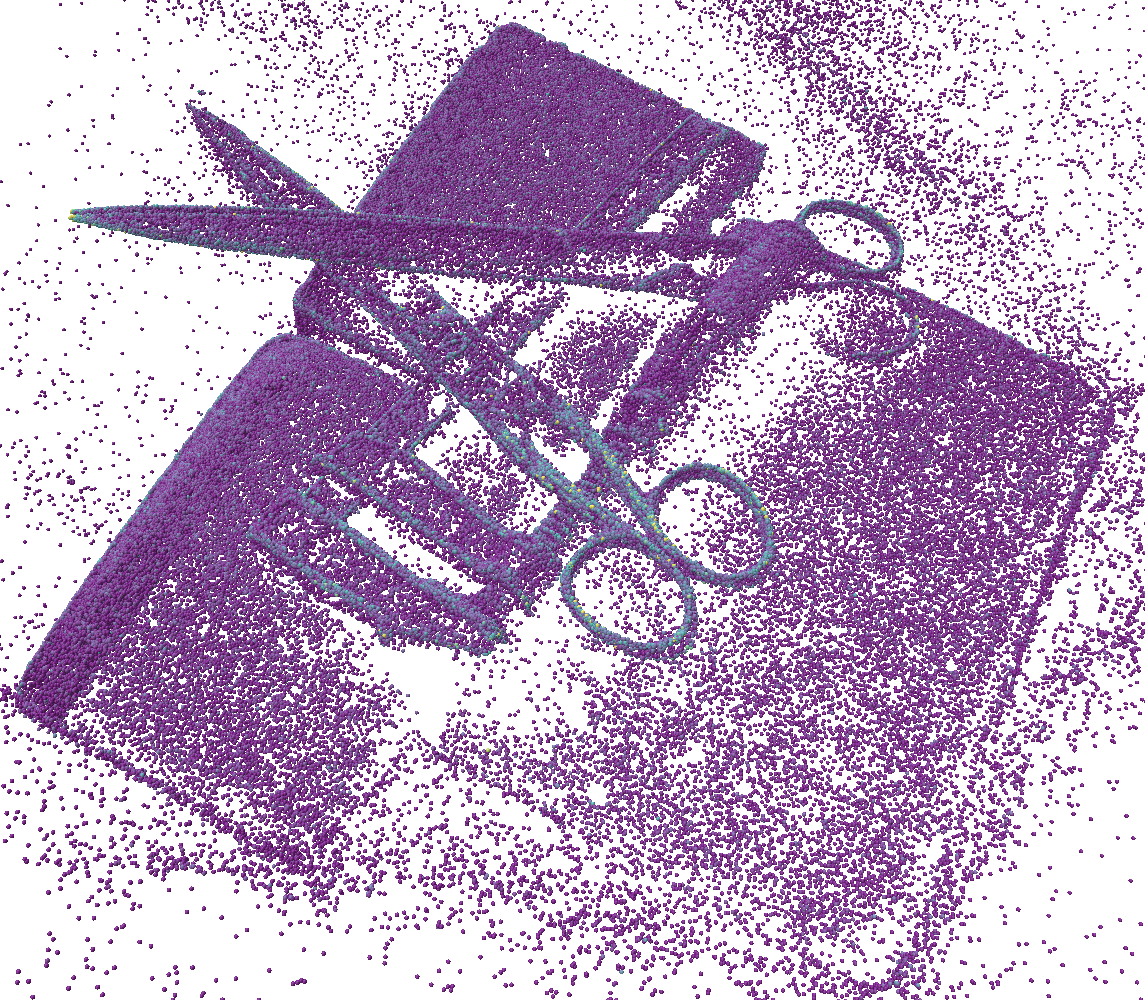} \\
        \includegraphics[width=0.24\linewidth]{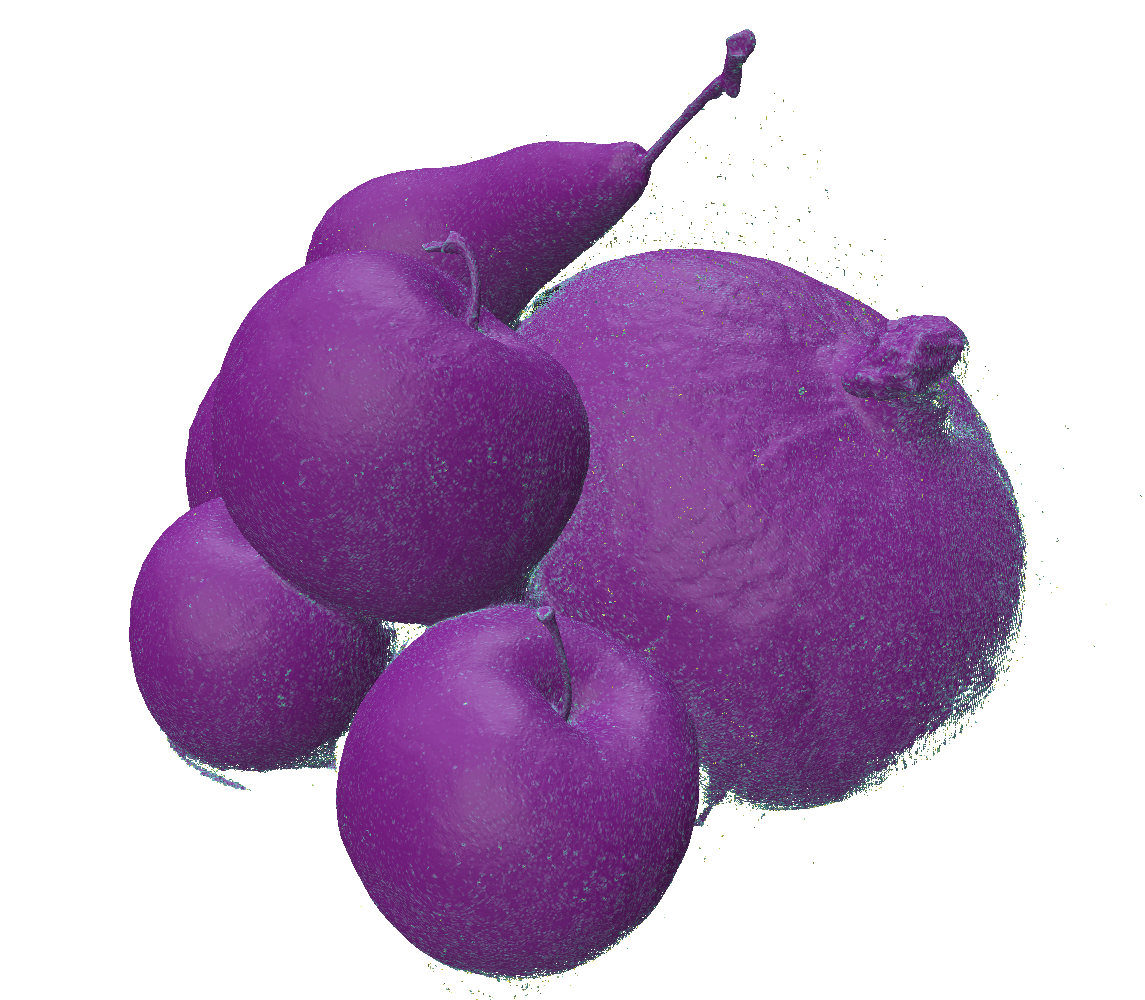} & \includegraphics[width=0.24\linewidth]{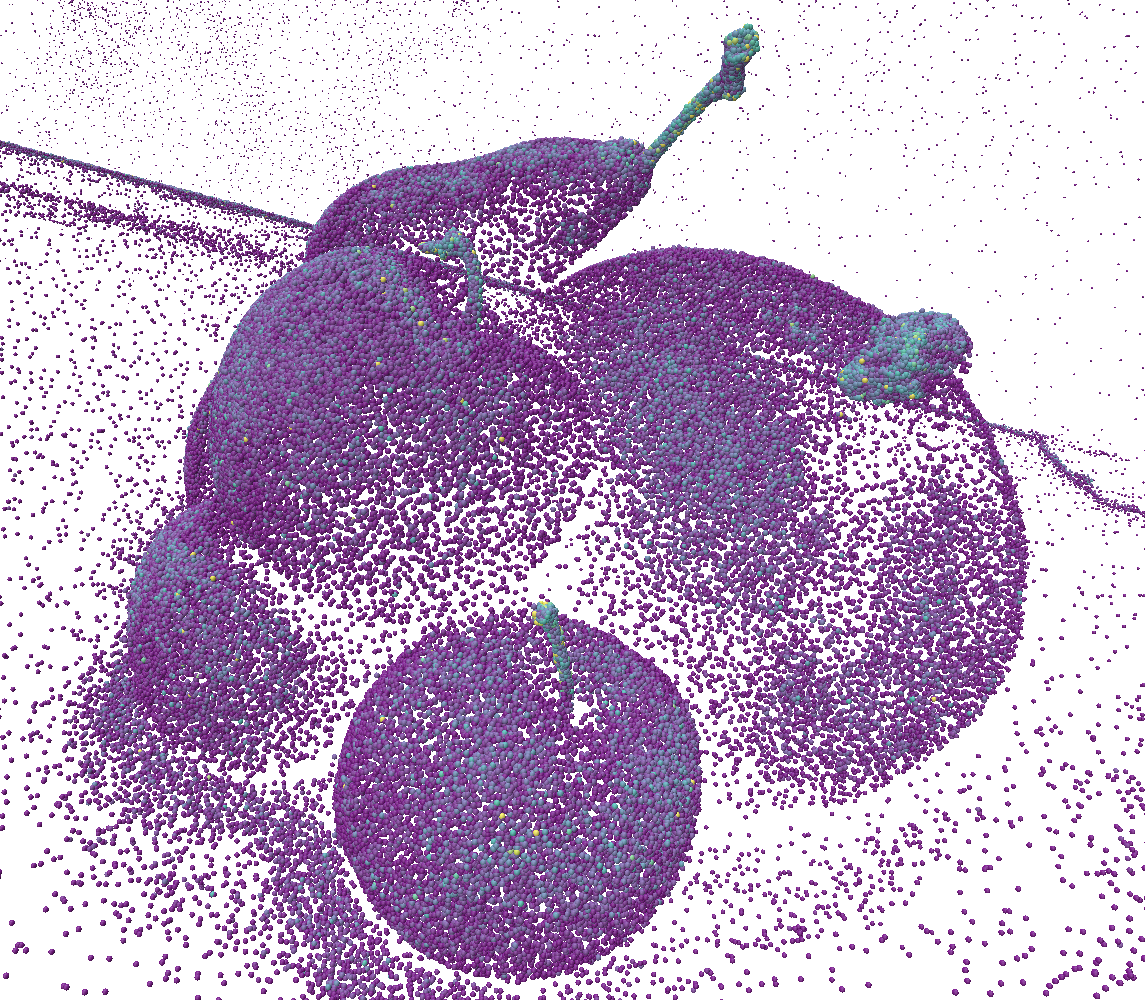} & \includegraphics[width=0.24\linewidth]{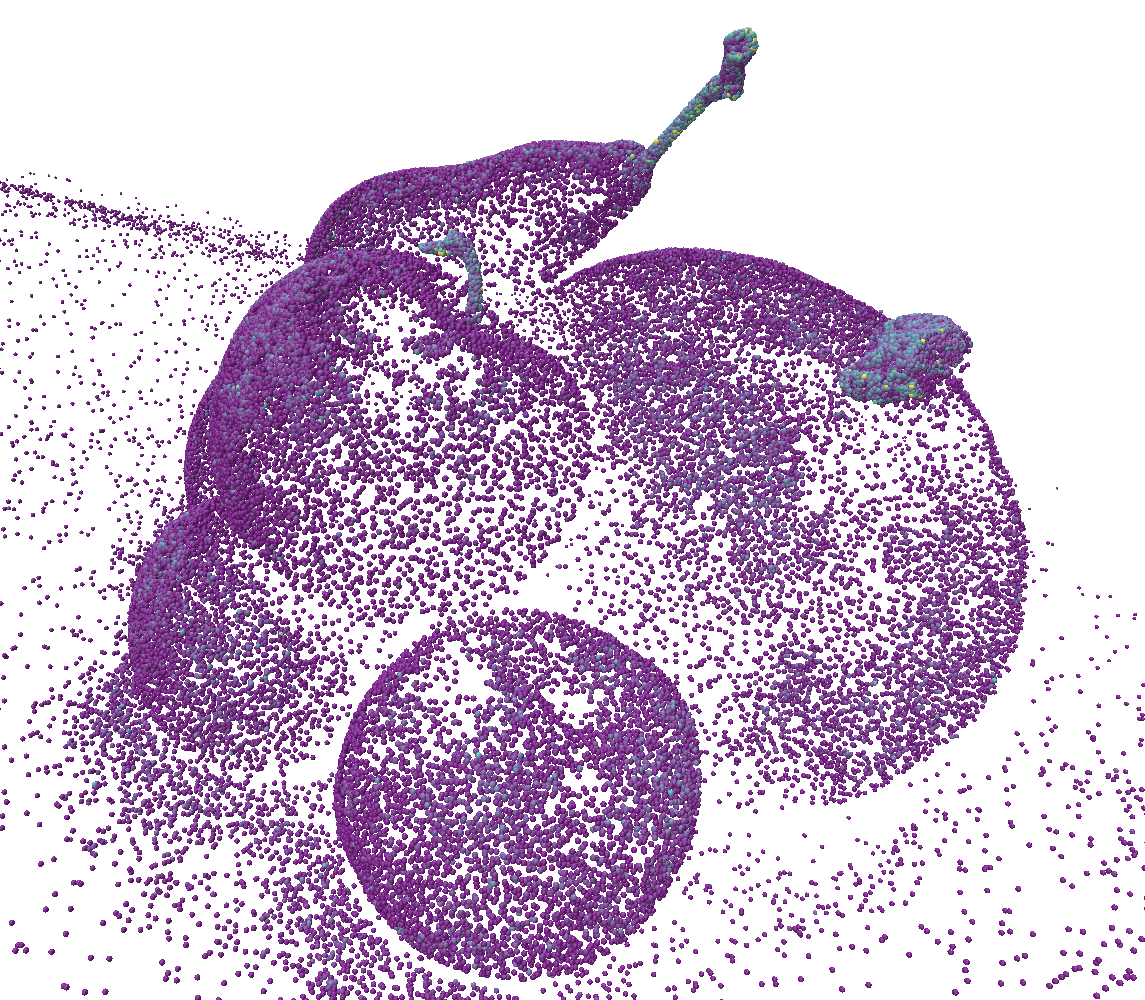} & \includegraphics[width=0.24\linewidth]{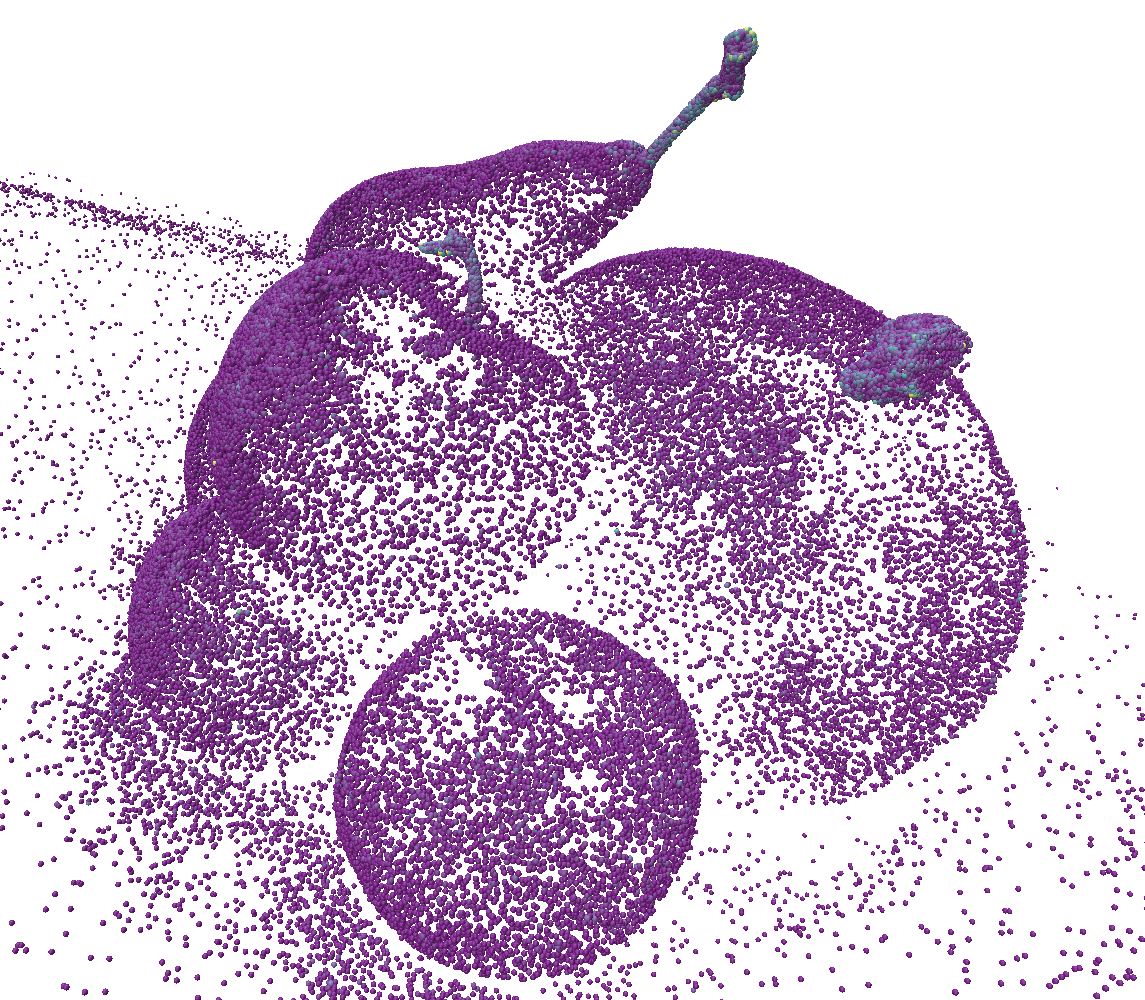}
    \end{tabular}

    \caption{Curvature computation using the Laplace-Beltrami operator of the mesh extracted from the 3DGS, the raw Gaussian center point cloud, our filtered result with Euclidean distance and our proposed method and filtering with Mahalanobis distance and normal direction on the DTU dataset (best view zoomed in). Purple indicates low curvature and turquoise to yellow higher curvature. (Top row) Our method does not overestimate curvature in areas with many outliers as the point cloud Laplacian or the pure Euclidean neighborhood does. (Bottom row) Even a clean-looking extracted mesh will add high errors close to mesh boundaries which our method does not suffer from. The resulting curvature function of our approach also represents the smooth surface better. }
    \label{fig:curvature}
\end{figure*}

We propose a novel framework that enables direct geometry processing on Gaussian splats without conversion into an intermediate representation, making the process lighter and more flexible. 
The two key aspects are: (1) a graph-based outlier removal strategy specifically designed for Gaussian distributions, and (2) an LBO formulation that leverages the Mahalanobis distance to incorporate covariance information as surface cues.
This approach is computationally efficient while fully utilizing the implicit geometric information encoded in 3D Gaussian splatting. 
Fig.~\ref{fig:curvature} shows the advantages on the application of curvature computation.

\paragraph{Contributions. } Our core contributions are as follows: 
\begin{enumerate}
    \item A definition of the Laplace-Beltrami operator that is computed directly on the Gaussian splatting representation and can fully leverage the geometric information encoded by the covariance. 
    \item An effective graph-based outlier removal method that is tailored for Gaussian distributions and preserves geometric structure while reducing noisy artifacts.
    \item We experimentally demonstrate the effectiveness of this LBO formulation on a variety of geometry processing applications, including geodesic distance computation and shape matching, showing clear improvement over mesh-based and point cloud-based baselines.
\end{enumerate}

\section{Related Work}
In this section, we show connections to the most relevant related work. A broader introduction into Gaussian splatting can be found in the survey of \cite{chen2024survey3dgaussiansplatting}.

\subsection{Gaussian Splatting}\label{sub:rw:gaussian}

3D Gaussian splatting was introduced in \cite{kerbl3Dgaussians} as an efficient framework for novel-view synthesis. 
It optimizes over a set of Gaussian distributions in 3D space with color and opacity values which can be rendered from new view points. 
While this works exceptionally well and has been applied to many applications~\cite{SplattingAvatar:CVPR2024,yan2024street}, the pipeline is focused on clean-looking rendering results but not clean geometry. 
To overcome this, several methods introduced additional constraints that focus on the geometric accuracy in the optimization. 
For example, the method of \cite{Huang2DGS2024} restricts the variance to a 2D plane (the dimension of the surface). 
Another approach, SuGaR~\cite{guedon2023sugar}, extracts a mesh and realigns the Gaussian splats with the surface of this mesh to obtain a cleaner geometry. 
Gaussian opacity fields~\cite{yu2024gaussian} proposes an approximation of the surface normals and incorporates normal regularization for better geometry. GSDF~\cite{yu2024gsdf} and 3DGSR~\cite{lyu20243dgsr} combine Gaussians with a signed distance function to further regularize the results with geometric losses. PGSR~\cite{chen2024pgsr} proposes a better estimation of the depth map and normal map, which allows single-view and multi-view regularization.
While these methods lead to better geometric accuracy, the extracted mesh is often noisy and/or overly smooth, and typically contains a large number of vertices.

\subsection{Laplace-Beltrami Operator}\label{sub:rw:lbo}

\paragraph{Laplacian Operator on Meshes.} 
The Laplace-Beltrami operator (LBO) is the generalization of the second derivative on general manifolds and a popular tool in many geometry processing applications. 
Its discretization, especially on triangular meshes, has been studied extensively and it has been shown that not all properties of the continuous Laplacian can be fulfilled in the discrete case at the same time~\cite{wardetzky2007nofreelunch}. 
While it is possible to use the graph Laplacian~\cite{taubin1995fairsurface} on a mesh by discarding the face information, this fails to take into account all information about the local geometry. 
More advanced mesh LBOs, like the cotan discretization~\cite{pinkallporthier,meyer2003discrete} or intrinsic Delaunay discretization~\cite{bobenko2007simplicial}, provide a more accurate approximation of the continuous behaviour. 
These can also be extended to more complex domains, like n-dimensional data~\cite{crane2019ndim}, general polygonal meshes~\cite{alexa2011polygonal,bunge2020polygon}, or non-manifold meshes~\cite{sharp2020nonmanifold}.
However, all of these depend on explicitly given connectivity information which can guarantee certain properties but does not work for less structured shape representations, like point clouds or Gaussian splats.

\paragraph{Laplacian Operator on Point Clouds.} 
As the discrete Laplacian operator relies on the definition of the neighborhood that is not explicitly given in the point cloud, it is essential to estimate the neighborhood function in a good way so that it approximate the intrinsic connectivity.
A straightforward solution would be to triangulate the point cloud, however, this is expensive and often leads to errors on sparse or noisy point clouds. 
Instead, a common solution is to \emph{locally} approximate the surface via its tangent plane and projection of surrounding points onto it, often by a nearest neighbor search~\cite{Belkin2009ConstructingLO}. 
This works well on smooth or flat regions, but struggles around very sharp features. 
The resulting inaccuracies can be diminished by building an operator that is robust to incorrectly found and non-manifold surface connections~\cite{sharp2020nonmanifold}, or by 
employing improvements in the surface estimation for these cases, for example through anisotropic Voronoi diagrams~\cite{Qin2018anisotropic}, or physic dynamics~\cite{petronetto2013meshfree}.
The recent work of \cite{pang2024neurallaplacianoperator3d} avoids direct estimation of the surface by learning the behavior of the LBO on different examples and then generalizing the behavior directly to new point clouds. 
While the centers of Gaussian splats do form a point cloud on which the previous methods can be applied, the directional variance at each point provides valuable additional information about the surface. 
In this work we propose a more accurate way to extract the LBO on Gaussian splats which include the variances in the surface estimation. 

\section{Background}
\label{sec:background}

\subsection{Gaussian Splatting} \label{sub:bg:gaussian}

3D Gaussian splatting~\cite{kerbl3Dgaussians} represents a scene as a collection of 3D Gaussian distributions, where each Gaussian is parametrized by $\{ (\mu_i, \Sigma_i, \alpha_i, c_i) \}_i$ with a mean $\mu_i \in \mathbb{R}^3$, covariance matrix $\Sigma_i \in \mathbb{R}^{3 \times 3}$, opacity $\alpha_i$, and color function $c_i$ encoded in spherical harmonics. 
Each Gaussian represents a local region of the scene with the mean defining its spatial location, the covariance matrix encoding its 3D shape and orientation, and the opacity and color function contributing to the final rendering.
This collection can be easily projected to 2D and rendered by accumulating density along a ray. 
The optimization process aims to match the render of these parameters to a collection of images from different view points which leads to bias towards photometric consistency instead of geometric accuracy, leading to Gaussians that may not be localized exactly on the object surface but still produce visually compelling results.

The rendering-focused optimization creates several challenges for geometry processing applications. First, many Gaussians may be positioned in free space or within object interiors if they contribute to visual quality. Second, low-opacity Gaussians that minimally affect rendering quality are retained in the representation, creating outliers for geometric analysis. 
Recent work has addressed these limitations through surface regularization using signed distance functions~\cite{guedon2023sugar}, depth and normal supervision~\cite{yu2024gaussian}, and constraining the Gaussian covariance to be 2-dimensional and aligned to a surface~\cite{Huang2DGS2024}.

\subsection{(Discrete) Laplace-Beltrami Operator} \label{sub:bg:lbo}

The Laplace-Beltrami operator (LBO) $L = \text{div}\cdot \nabla$ generalizes the second derivative to general closed compact manifolds. 
This operator, along with its eigenfunctions $\phi_i$ and eigenvalues $\lambda_i$ which are non-trivial solutions to $L \phi_i = \lambda_i \phi_i$, are popular tools in geometry processing, enabling applications like spectral analysis and functional maps (see \cref{sub:rw:lbo}). 

When discretizing the underlying manifold, for example as a mesh or point cloud, the continuous LBO must be approximated as well which leads to discretization artifacts \cite{wardetzky2007nofreelunch}. 
For the triangular meshes, the cotan discretization~\cite{pinkallporthier} is the most widely used version. It defines the stiffness matrix as
\begin{align}
    W_{ij} = \begin{cases}
        \frac{1}{2} (\text{cot}\alpha_{ij} + \text{cot} \beta_{ij}), & \text{if } (i,j) \in E \\
        - \sum_{k \in \mathcal{N}(i)} w_{ik}, & \text{if } j=i \\
        0, & \text{otherwise}
    \end{cases}\label{eq:cotanlbo}
\end{align}
where $\alpha_{ij}, \beta_{ij}$ are the opposing angles to the edge between vertices $v_i, v_j$ 
(see \cref{fig:mahalanobis}) and $\mathcal{N}(i)$ the set of vertices in the 1-neighborhood of vertex $i$. 
The discrete LBO can the be computed as $L = M^{-1} W$, where $M$ is a diagonal mass matrix, describing the local weight at each vertex $M_{ii}$ as the area of the voronoi cell around vertex $i$.
This discretization of the Laplacian does not necessarily fulfill the maximum principle~\cite{wardetzky2007nofreelunch}, but will if applied the intrinsic Delaunay triangulation of the mesh \cite{bobenko2007simplicial}.
However, this can only be computed clean, pure triangular meshes without non-manifoldness (e.g., edges with three triangles attached). 
For meshes of arbitrary topology, \cite{sharp2020nonmanifold} suggested to use the tufted cover, which generates an implicit manifold overlay of a given connectivity, in combination with the intrinsic triangulation. 
Due to its flexibility with respect to the connectivity, the tufted Laplacian is well-suited to be used on point clouds by approximating a local neighborhood and connectivity for each point without the need to reconstruct a full, consistent triangle mesh, which is expensive and prone to be noisy. 
We will use this property but extend the neighborhood definition to be explicitly designed for Gaussian distributions. 
\section{Method}
\label{sec:method}

We propose a discretization of the Laplace-Beltrami operator (LBO) that operates directly on 3D Gaussian splatting representations, preserving encoded geometric information without requiring a conversion to an intermediate format. 
Our approach addresses two fundamental challenges: constructing an appropriate local neighborhood in the absence of explicit connectivity information, and handling noisy outliers that arise from rendering-focused optimization.
First, we introduce a graph-based filtering procedure that leverages symmetric Mahalanobis neighborhoods to remove geometric outliers while preserving surface structure, see \cref{sub:method:graph}.
Second, we formulate an Laplace-Beltrami operator that is designed to use the full covariance information of the Gaussians to capture local geometric relationships, see \cref{sub:method:laplacian}.

\subsection{Graph-based Outlier Removal}\label{sub:method:graph}
3D Gaussian splatting optimization often generates outliers that only contribute minimally to the rendering quality but significantly degrade geometric accuracy. 
While the original 3DGS~\cite{kerbl3Dgaussians} already proposed to remove near-zero opacity splats during training, many low-information Gaussians persist inside the objects and close the surface~\cite{yu2024gaussian}.
Due to their low opacity or position inside of closed objects, these splats can take high variance forms in all direction without changing the rendering result. 
We propose to use a symmetric Mahalanobis neighborhood construction that can detect such outliers.

\paragraph{Mahalanobis Distance.} 
The Mahalanobis distance is a measure of the distance between a point $p \in \mathbb{R}^d$ and a distribution $\mathcal{G}$ with mean $\mu \in \mathbb{R}^d$ and covariance $\Sigma \in \mathbb{R}^{d \times d}$.
We only consider $d=3$ in this paper.
It is defined as
$$d_{M}(p, \mathcal{G})=\sqrt{(p-\mu)^T\Sigma^{-1}(p-\mu)}.$$ 
See \cref{fig:mahalanobis} for a visualization and relation to Euclidean distance.
The metric captures the anisotropic structure of Gaussians, where surface-aligned splats typically exhibit high variance along tangent directions and low variance in normal direction.
Thus, a neighborhood based on the Mahalanobis distance clusters together points that form a coherent surface while assigning outliers a higher distance. 

However, we can only compute the distance between a \emph{point} and a distribution. What we actually want to do is compute the distance between two Gaussian distributions. 
Computing the distance between two distributions is possible, for example, with the earth mover's distance, but computationally very expensive. 
In practice, we approximate by taking a symmetric approach, computing the Mahalanobis distance between center points and surrounding Gaussian distributions, and building neighborhoods between splats where the center points are both each other's Mahalanobis neighbors.

\begin{figure}
    \centering
    \begin{tikzpicture}[scale=0.8]
    \coordinate (A) at (0,0);
    \coordinate (B) at (1.3,0.8);
    \coordinate (C) at (-0.8,1.3);

    \foreach\i in {0.2,0.4,...,1} {
        \fill[opacity=\i,BlueGreen,rotate around={30:(A)}] (A) ellipse ({2-2*\i} and {2-2*\i});         
      }

    \draw[dashed] (A) -- (B) node [pos=.7, label=below:$d_E$] {};
    \draw[dashed] (A) -- (C) node [pos=.5, label=left:$d_E$] {};

    \draw[fill=black, draw=black] (A) circle (2pt);
    \draw[fill=black, draw=black] (B) circle (2pt);
    \draw[fill=black, draw=black] (C) circle (2pt);

    \node (p) at (1.6,0.8) {p};
    \node (q) at (-1.1,1.3) {q};
    \node (G) at (-0.9,-0.5) {$\mathcal{G}_1$};
\end{tikzpicture}%
 \quad\quad \begin{tikzpicture}[scale=0.85]
    \coordinate (A) at (0,0);
    \coordinate (B) at (0.7,0.3);
    \coordinate (C) at (-0.8,1.3);

    \foreach\i in {0.2,0.4,...,1} {
        \fill[opacity=\i,BlueGreen,rotate around={30:(A)}] (A) ellipse ({1-\i} and {2-2*\i});         
      }

    \draw[dashed] (A) -- (B) node [pos=.7, label=below:$d_M$] {};
    \draw[dashed] (A) -- (C) node [pos=.5, label=left:$d_M$] {};

    \draw[fill=black, draw=black] (A) circle (2pt);
    \draw[fill=black, draw=black] (B) circle (2pt);
    \draw[fill=black, draw=black] (C) circle (2pt);

    \node (p) at (1,0.3) {p};
    \node (q) at (-1.1,1.3) {q};
    \node (G) at (-0.7,-0.5) {$\mathcal{G}_2$};
\end{tikzpicture}
    \caption{
    Difference between the Euclidean distance (left) and Mahalanobis distance (right). $p$ and $q$ both have the same distance to distribution $\mathcal{G}_i$. The Euclidean distance is a special case of the Mahalanobis distance if the covariance is equal in all directions while the general Mahalanobis distance weights directions differently based on the directional variance. }
    \label{fig:mahalanobis}
\end{figure}

\paragraph{Graph Construction}

\begin{figure}
    \centering
    \includegraphics[width=0.32\linewidth]{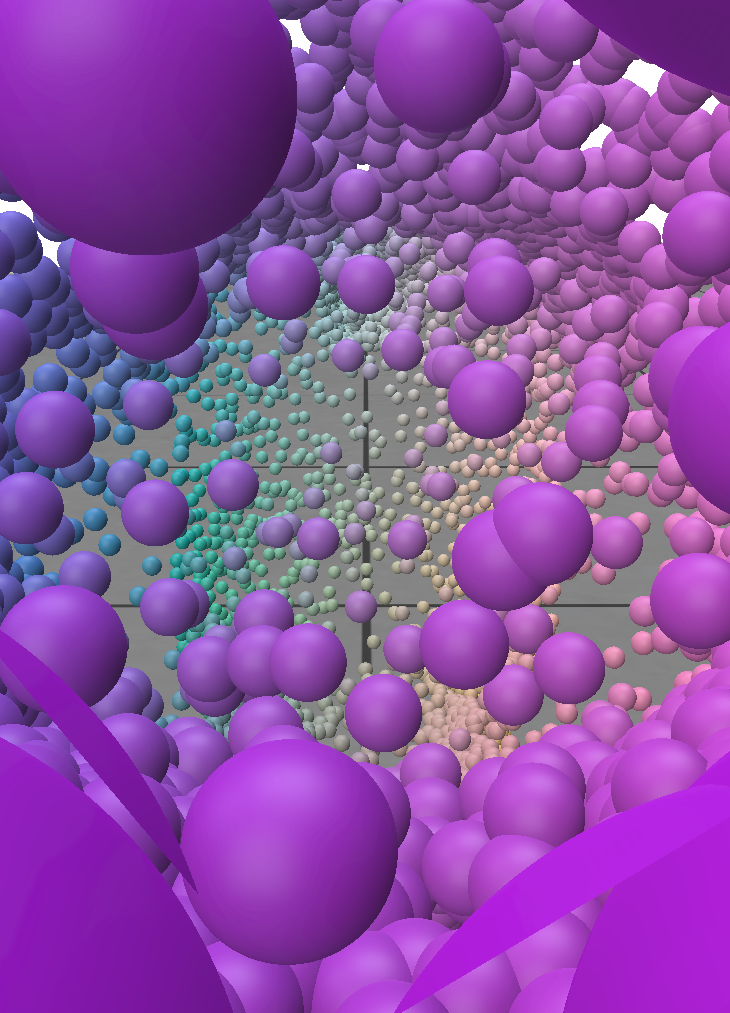}
    \includegraphics[width=0.32\linewidth]{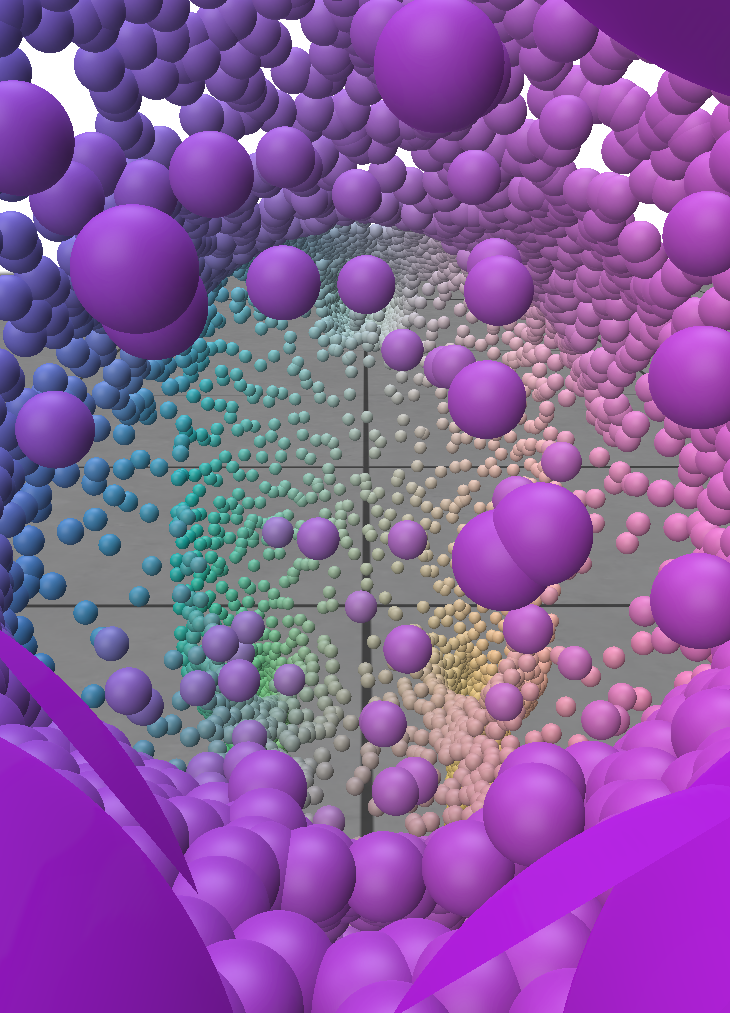}
    \includegraphics[width=0.32\linewidth]{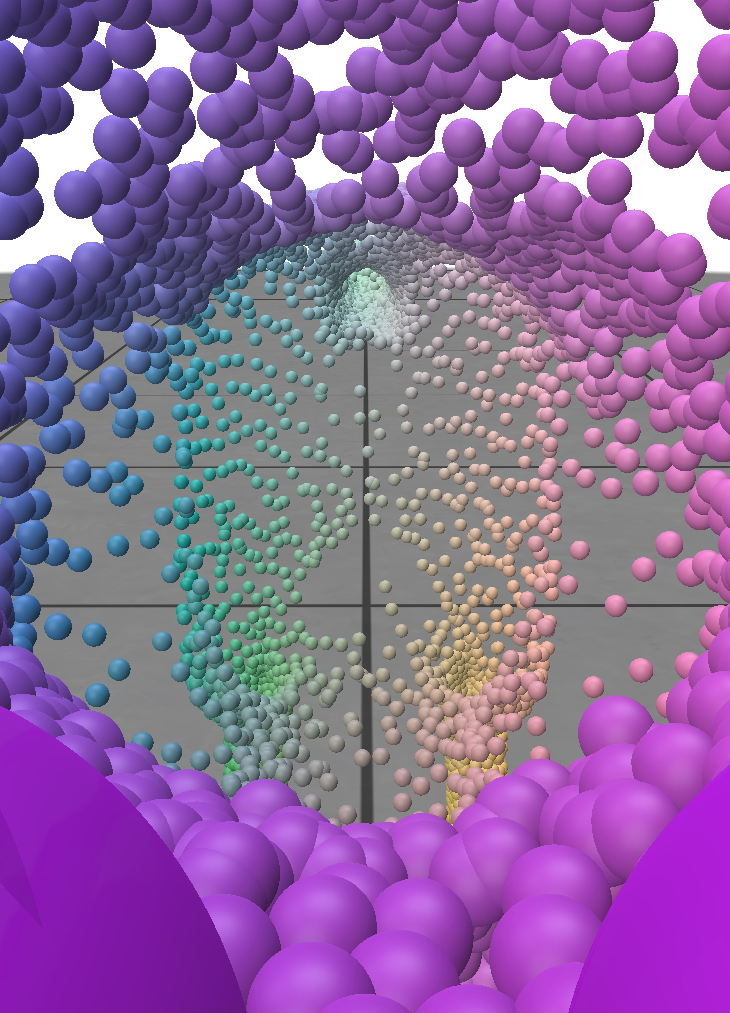}
    \caption{Example of the inside of the point cloud before (left), after (middle) cleaning using the training from GOF~\cite{yu2024gaussian} and using adaptive training (right).}
    \label{fig:filtering}
\end{figure}

We compute the following neighborhood graph for each Gaussian center:
$(p_i, p_j)$ is an edge if and only if $p_i$ is among the $k$-nearest Mahalanobis neighbors of $p_j$ \textbf{and} $p_j$ is among the $k$-nearest Mahalanobis neighbors of $p_i$. 
This bidirectional constraint leads to a robust surface connectivity while naturally isolating outliers. 
The resulting graph typically contains multiple connected components due to interior outliers and noise.
We choose only the largest connected component and prune all Gaussians belonging to others.
This filtering leads to a sparser and cleaner representation of the geometry,
see \cref{fig:filtering} for a visualization of the effect.
In addition, \cref{sub:exp:adapted} shows experimentally that this efficient processing during training leads to a very compact and clean representation.
For larger scenes, that are expected to have multiple disconnected but relevant components, it is possible to only prune components below a threshold size.

\subsection{Gaussian Laplacian}\label{sub:method:laplacian}
The LBO encodes local connectivity and geometry. 
Gaussian splats do not encode connectivity explicitly but the covariance implies the surface geometry. 
The common point cloud Laplacian~\cite{sharp2020nonmanifold} approximates the surface by using Euclidean neighborhoods of each point. 
We extend this idea to leverage the full geometric information encoded in the Gaussian covariance matrices. 
Instead of Euclidean distance, we construct $k$-nearest neighbor relationships using the Mahalanobis distance, enabling the neighborhood structure to respect the anisotropic geometry encoded in each Gaussian. 
This approach naturally weights connections based on both spatial proximity and geometric compatibility.

We adapt the robust point cloud Laplacian formulation to operate on our cleaned Gaussian centers with Mahalanobis-based connectivity. 
The resulting operator inherits the mathematical properties of the non-manifold and point cloud Laplacian~\cite{sharp2020nonmanifold} while incorporating the geometric information of Gaussian covariance matrices.
\cref{sec:experiments} experimentally validates these changes. 

\paragraph*{Normal Estimation. }
The non-manifold Laplacian~\cite{sharp2020nonmanifold} contains a normal estimation step from the point neighborhood.
This estimation is strongly effected by outliers and noise and for the Gaussian Laplacian we replace it by estimating the normal of each point as the eigenvector corresponding to the smallest eigenvalue of the covariance matrix of the corresponding Gaussian distribution.
This change is applicable under the assumption that the Gaussian splats are mostly flat and well aligned to the surface, as is the case in methods such as SuGaR~\cite{guedon2023sugar}, GOF~\cite{yu2024gaussian} and PGSR~\cite{chen2024pgsr}.

\paragraph*{Adaptive Training of 3DGS}
We integrate our outlier removal directly into the 3DGS training process to improve geometric quality.
Specifically, we filter the Gaussian splats every $N=3000$ iterations by removing all except the $k=1$ biggest components which prunes all outliers.
$k$ can be increased for larger scenes.
Note that finding all connected components can be achieved by breadth-first search in complexity $O(N)$, hence the extra cost of computation is not significant during training.

\begin{figure*}[ht]
    \centering
    \includegraphics[width=0.7\linewidth]{tikz_figures/eigval_mahalanobis.tex}
    \caption{Statistical analysis over all the objects on the loss of eigenvalues. The index represents the order of the eigenvalue by magnitude. The line represent the average of the loss and one std is used for the confidence interval. Our method outperforms the Mesh~(GOF). }
    \label{fig:eigval}
\end{figure*}
\begin{table*}[htb]
\centering
   \resizebox{.7\linewidth}{!}{%
\begin{tabular}{c|c||c|c|c|c|c}
\hline
     Category & Mesh (GT) & Point Cloud & Mesh (GOF) & Ours (Euclid) & Ours (M+N) & Ours (AT+M+N)\\ \hline
     Cat & 0.017     & 0.045 &  0.028 &0.034  & \lightbold{0.027} & \bf{0.026} \\
     Centaur & 0.018 & 0.036 &  0.024 & \lightbold{0.023}  & \lightbold{0.023} & \bf{0.022} \\
     David & 0.036   & 0.050 & 0.058 & \lightbold{0.038}  & \bf{0.035} & \bf{0.035} \\
     Dog & 0.016     & 0.050 & 0.037 & 0.036  & \bf{0.031} & \lightbold{0.032} \\
     Horse & 0.016   & 0.046 & 0.029 & \bf{0.022}  & \bf{0.022} & \bf{0.022}\\
     Michael & 0.038 & 0.051 &  0.065 & \bf{0.046}  & \lightbold{0.047} & \bf{0.046} \\
     Victoria & 0.035& 0.117 &  \bf{0.078} &0.113  &0.095 & \lightbold{0.082}\\
     Wolf & 0.012    & 0.036 &  \bf{0.010} & \lightbold{0.017}  &0.018 & 0.018 \\
\hline
    Total & 0.025 & 0.056 & 0.045 & 0.045 & \lightbold{0.040} & \bf{0.038} \\ \hline
\end{tabular}%
}
\caption{Average error in geodesic computation $\mathcal{E}_{geo}$ by Geodesics in Heat~\cite{crane2017heat} in comparison to the exact distance on the ground truth mesh. The bold represents the best score and the gray represents the second best. Note that Mesh (GT) represents the approximation error introduced by Geodesics in Heat on the ground-truth mesh as a reference and is not a competitor as the ground truth is not known in general.}
\label{table:heat}
\end{table*}

\begin{table*}[h]
\centering
\begin{tabular}{c c c c c c}
\hline
Dataset & Mesh~(GT) & 3DGS & Mesh~(reconstructed) & Ours (Outlier Removal) & Ours (Adaptive Training) \\ \hline
    TOSCA & 30416 & 19394 & 197364 & 17763 & 14523 \\

Shiny Blender & {\centering /} & 169877 & 3546714 & {\centering /} & 99453\\ \hline
\end{tabular}%
\caption{Comparison of the average size of different representations. For meshes we report the number of vertices. For 3DGS we report the number of splats. We use GOF to obtain the 3DGS and reconstructed mesh on TOSCA, and use PGSR on Shiny Blender. The number of vertices in the reconstructed mesh is ten times more than that of 3D Gaussian splatting while still performing worse in several quantitative evaluations, see \cref{sec:experiments}. }
\label{table:size}
\end{table*}

\begin{table*}[h]
\centering
\begin{tabular}{c c c c c c c}
\hline
Method &  Car & Coffee & Helmet & Teapot & Toaster & Mean \\ \hline
     3DGS~(PGSR) &    26.43 & 30.86 & 25.72 & 37.08 & 20.24 & 28.07\\
     Ours (Adaptive Training) & 26.21 & 30.82 & 25.67 & 36.96 & 20.04 & 27.94\\
\hline
\end{tabular}%
\caption{Comparison of the rendering quality with and without adaptive training on the Shiny Blender dataset~\cite{verbin2022ref} measured by PSNR. In the end, adaptive training does not notably degrade the rendering quality but is able to remove a significant amount of outliers. }
\label{table:rendering}
\end{table*}

\begin{figure*}[htb]
    \centering 
    \begin{overpic}[width=0.8\linewidth]{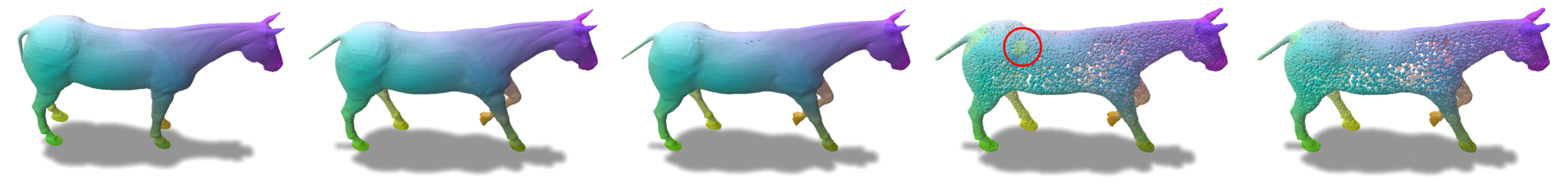}
        \put(3,-2){\colorbox{white}{\footnotesize{Source}}}
        \put(26,-2){\colorbox{white}{\footnotesize{Mesh (GT)}}}
        \put(42,-2){\colorbox{white}{\footnotesize{Mesh (GOF)}}}
        \put(66,-2){\colorbox{white}{\footnotesize{PC}}}
        \put(83,-2){\colorbox{white}{\footnotesize{Ours(M+N)}}}
    \end{overpic}
    \caption{Color map for shape matching in the horse category. 
    The source map (center) is colored by coordinate and the colors transferred with the correspondence computed by each method to the surrounding shapes.
    Inconsistencies in color indicate problems with the correspondence.
    }
    \label{fig:cmap:shape_matching}
\end{figure*}

\section{Experiments}
\label{sec:experiments}

We show the effectiveness of our method through comprehensive evaluation against established baselines. 
Our Gaussian Laplacian is compared against the robust point cloud Laplacian~\cite{sharp2020nonmanifold} as a baseline and the cotan Laplacian~\cite{pinkallporthier} on meshes, either on the ground-truth mesh if it exists or on meshes extracted from 3DGS results if it does not exist.
Since common 3DGS datasets lack ground-truth meshes due to their focus on rendering, they are not suitable for quantitative evaluation. Instead, we turn to the datasets that are commonly used in geometry processing for quantitative evaluation, see \cref{sub:exp:evaluation}.
Our experiments demonstrate superior performance when computing heat diffusion and geodesics in heat (\cref{sub:exp:heatandgeo}), competitive results in shape correspondence (\cref{sub:exp:matching}), and significant improvements in the compactness and cleanliness of 3DGS representations when our outlier detection was integrated in the training process (\cref{sub:exp:adapted}). 
Gaussian splatting representations are generated using either \cite{yu2024gaussian} or \cite{chen2024pgsr} for all experiments, depending which works better on each dataset.
All experiments were done with an NVIDIA RTX 4090 with 24~GB~RAM. 

\subsection{Evaluations and Comparisons}\label{sub:exp:evaluation}

\paragraph{Dataset.}
We generated a dataset for evaluation from TOSCA~\cite{MB08}, which provides $8$ classes of non-rigidly deformed meshes with ground-truth correspondence and at least 3 different poses in each class.
Since common 3DGS datasets contain neither ground-truth meshes nor correspondences, we use this to provide reliable quantitative evaluations.
We used Blender to render an image dataset of TOSCA suitable for 3DGS optimization, generating 100 training images, 50 validation images, and 30 test images per shape across 5 poses per class (or maximum available). 
Objects are scaled by $\frac{1}{20}$ to fit into a default camera frame.
We also experiment on Shiny Blender~\cite{verbin2022ref}, an object-centric dataset that contains test views for the evaluation of novel view rendering with reflective surfaces, to show the effectiveness of the adaptive training.
Qualitative results on data commonly used for 3DGS (DTU~\cite{jensen2014large} and TnT~\cite{knapitsch2017tanks}) can be found in \cref{fig:curvature} and the supplementary.

\paragraph{Evaluation Metrics.}

Since the LBO matrices across different 3D representations cannot be compared directly due to varying discretization, e.g. ordering and amount of vertices, we evaluate derived properties of the operators to compare performance. 
Derived point-wise functions, like curvature, can not be compared directly under varying discretizations, either. 
In that case, we project each point to its Euclidean nearest neighbor on the ground-truth mesh and compare with this value. 
The evaluation is done with the eigenvalues, a curvature loss, the geodesic distances and shape matching accuracy.

\paragraph{Competing Methods. }
We compare our method with the following baselines since no other direct approaches that define the LBO for Gaussian splatting representations exist:
\begin{itemize}
    \item \textbf{Point Cloud (PC):} the robust point cloud Laplacian computed on centers of the 3D Gaussian splats~\cite{sharp2020nonmanifold},
    \item \textbf{Mesh~(GOF):} the tufted Laplacian~\cite{sharp2020nonmanifold} on a mesh reconstructed from the state of the art 3DGS approach Gaussian opacity fields~\cite{yu2024gaussian},
    \item \textbf{Ours~(Euclid):} our proposed Gaussian Laplacian using the k-nearest Euclidean neighbors to compute the normal direction (as in \cite{sharp2020nonmanifold}), applied on the outlier-filtered 3DGS (very similar to the robust point cloud Laplacian on the outlier-filtered point cloud),
    \item \textbf{Ours~(M+N):} our proposed Gaussian Laplacian using the  k-nearest Mahalanobis neighbors and lowest magnitude variance direction as the normal, applied on the outlier-filtered 3DGS (\cref{sub:method:graph}),
    \item \textbf{Ours~(AT+M+N):} using the  k-nearest Mahalanobis neighbors and lowest magnitude variance direction as the normal, applied on the 3DGS optimized by our proposed adaptive training (\cref{sub:method:laplacian}).
\end{itemize}
If given, we compute the Laplace-Beltrami operator on the ground-truth meshes as the optimal solution, and we use the state-of-the-art tufted Laplacian~\cite{sharp2020nonmanifold} to compute the Laplacian operator on both point clouds, meshes and our adaption. 

\subsection{Experimental Results}

We present a series of experiments showing the performance of our proposed Gaussian Laplacian in comparison to other baselines approximating the LBO on 3DGS. 

\paragraph{Eigenvalues. }
Following the convergence analysis of Belkin and Niyogi~\cite{belkin2006convergence} for the spectrum of the LBO, we  compare the eigenvalues of the normalized Laplacian matrix by solving $Lx = \lambda Mx$ and computing the first $K$ eigenvalues (the lower the more stable). 
Since the eigenvalue scales inversely proportional to surface area, we calculate the normalized difference $S|\lambda_i - \lambda_i^{gt}|$ for $i = 1,\ldots, 100$ for comparison. 

We take the first $K=100$ eigenvalues of each Laplacian operator and compute the difference to the operator on the ground-truth mesh. 
A shown in~\cref{fig:eigval}, our method and Mesh~(GOF) perform on-par, and both outperform the point cloud-based Laplacian. 
For the first 20 eigenvalues, which should be the most stable, our method is more accurate than even the mesh-based Laplacian and, compared to using Euclidean distance, the Mahalanobis distance makes the spectrum of the Laplacian operator closer to the ground truth.

\paragraph{Curvature}\label{sub:exp:curvature}
As an intrinsic property, mean curvature can be derived from the Laplace-Beltrami operator via the equation $\Delta p = -2Hn$, where $p$ denotes the point coordinates, $H$ the mean curvature and $n$ the unit normal vector. 
\cref{tab:mean_curvature} demonstrates that our Gaussian Laplacian with adaptive training estimates curvature much closer to the ground truth (based on the ground-truth mesh) than all other alternatives. 
Surprisingly, the curvature error is largest for the extracted mesh. 
This is due to artifacts produced in the extraction process to which curvature computation is highly sensitive. 
Behavior like this indicates that avoiding unnecessary representation conversion is not always a good choice and using representation-native operations has great merit. 
\begin{table}[hb]
    \centering
    \footnotesize
    \begin{tabular}{c|c|c|c|c}
    \hline
       Type  &  Mesh~(GOF) & PC & Ours~(M+N) & Ours~(AT+M+N)\\
       \hline
        Avg. & 43.049 & 17.746 & 16.317 & \bf{16.023} \\
        Min.  & 12.720 & 7.954 & 6.332 & \bf{6.202} \\
        Max.  & 129.369 & 33.621 & 29.867 & \bf{29.306}\\
        \hline
    \end{tabular}
    \caption{$L_1$-error of the mean curvature computed from the Laplacian operator on an extracted mesh, point cloud, ours with Mahalanobis distance and ours with adaptive training. }
    \label{tab:mean_curvature}
\end{table}

\paragraph{Geodesic Distance}\label{sub:exp:heatandgeo}
The geodesic distance describes the length of the shortest path between two points on a surface. 
It can be efficiently approximated using Geodesics in Heat~\cite{crane2017heat} which is based on heat diffusion, a function describing how heat diffuses over time that can be computed using the Laplacian operator.
As ground truth, we use an exact algorithm~\cite{mitchell1987discrete} for geodesic distances on the ground-truth meshes and compare the approximation quality of each method to this function. 

\begin{figure}[h]
    \centering
    \includegraphics[width=0.24\linewidth]{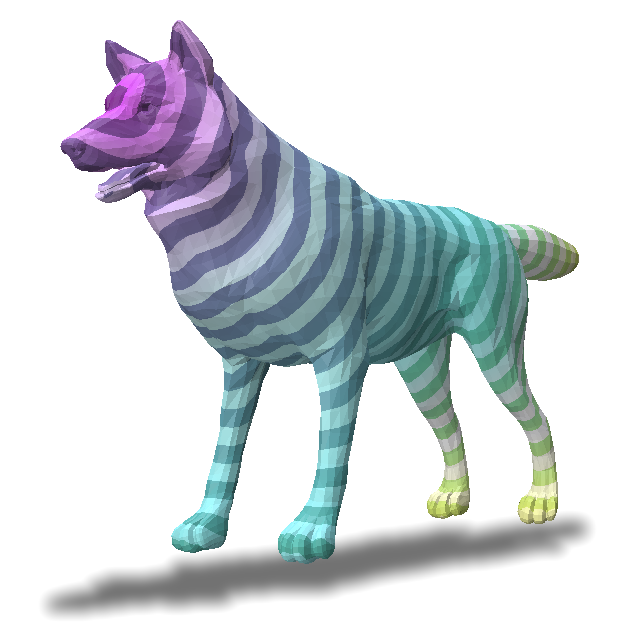}
    \includegraphics[width=0.24\linewidth]{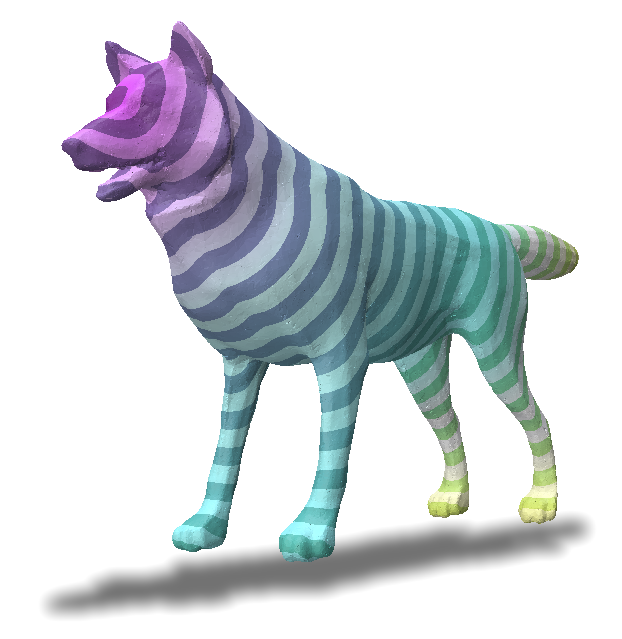}
    \begin{overpic}[width=0.24\linewidth]{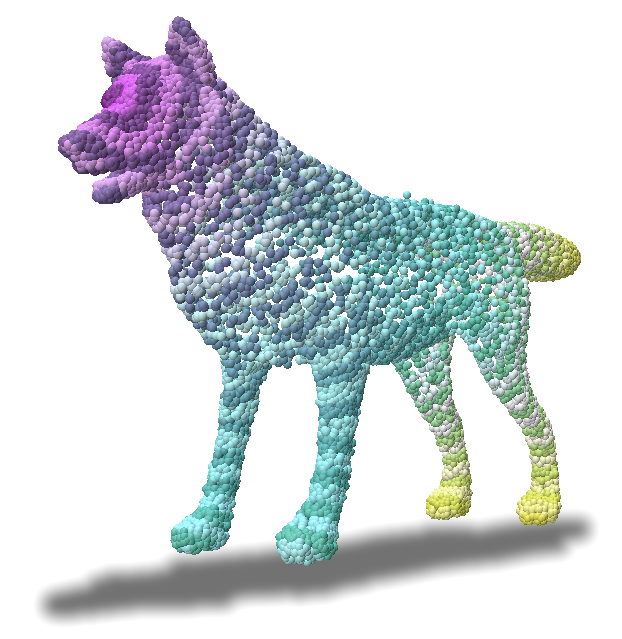}
    \put(46,15){\color{red}\circle{18}}
    \put(15,75){\color{red}\circle{18}}
    \end{overpic}
    \includegraphics[width=0.24\linewidth]{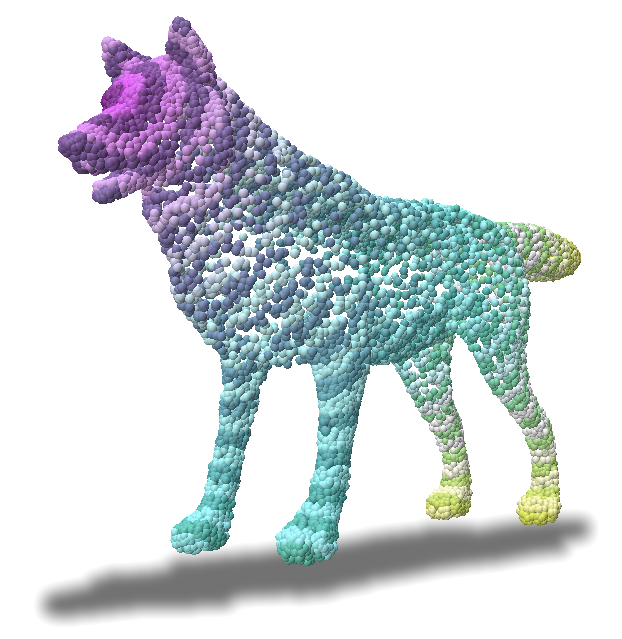}
    \caption{Geodesics in Heat~\cite{crane2017heat} computed on (from left to right) the ground-truth mesh, the reconstructed mesh (GOF), the center point cloud (uncleaned) and the cleaned Gaussian splats (ours). The cleaning significantly improves the results (see red circles). }
    \label{fig:heat_qualitative}
\end{figure}

Since computing exact geodesic distances is expensive, we uniformly sample $100$ source points on the ground-truth mesh and use only the distance functions of these points. 
We compute the error normalized by surface area:
$$\mathcal{E}_{geo}(\Delta) = \frac{1}{100}\frac{1}{\sqrt{S}}\sum_{i=1}^{100} \sum_{j=1}^N|d^j - d_{\Delta}^j|$$ 
where $S$ is the surface area, $N$ is the number of points of the ground-truth mesh, $d^j$ is the exact geodesic distance~\cite{mitchell1987discrete} from the point $j$ to the source point and $d_{\Delta}^j$ is the computed geodesic distance from $j$ to the source point using the LBO $\Delta$.
\cref{table:heat} shows the comparison of geodesic distance errors averaged over all $100$ distance functions. 
The Mesh~(GT) column is given as a reference to see the error induced by Geodesics in Heat algorithm to the exact geodesics on the ground-truth mesh, not as a direct competitor. 
Our method significantly outperforms the point cloud Laplacian in all cases and outperforms the extracted mesh-based computation in almost all cases.
In cases where the adaptive training does not improve the results, it is still on-par with the best results. \cref{fig:heat_qualitative} shows qualitative results.

\begin{figure}[h]
    \centering
    \includegraphics[width=1.0\linewidth]{tikz_figures/shape_matching_quant_mahalanobis_3dv.tex} 
    \caption{Cumulative geodesic error curve after representing the ground-truth correspondence in a low-dimensional functional map matrix, averaged over all the categories. 
    The x-axis shows the percentage of shape diameter and y-axis percentage of matches with lower error. Dashed only for visibility. 
    }
    \label{fig:quant:shape_matching}
\end{figure}

\subsection{Shape Matching}\label{sub:exp:matching}
A fundamental problem in geometry processing is finding correspondences between pairs of isometrically deformed 3D shapes. 
We evaluate correspondence quality in the functional maps framework~\cite{ovsjanikov2012functional}, which uses the Laplace-Beltrami eigenfunctions for dimensionality reduction of the problem, by testing preservation of the ground-truth correspondence after the representation change. 

Given the ground-truth correspondence as a permutation $P$, the functional map representation is computed as
\begin{equation}
   C = \Phi_t^\top M_t P \Phi_s
\end{equation}
where $\Phi_s, \Phi_t$ are the metric of the first $k$ (ordered by frequency) stacked eigenvectors of the source and target shape, respectively, and $M_t$ the mass matrix on the target shape. 
The point-wise correspondence $Q$ can be extracted again via nearest neighbors in the aligned spectral space between the row-wise entries of $\Phi_t$ and $C\Phi_s$, see \cite{ovsjanikov2012functional} for detailed explanation.
Since we do not want to evaluate a specific matching algorithm, this projection of the ground truth, which still leads to errors due to the dimension reduction, gives us a measure of quality for the eigenbasis and, in turn, our Laplace-Beltrami operator.

Using $100$ eigenfunctions on $1000$ uniformly sampled vertices, we compute the geodesic error of each Laplacian on the TOSCA dataset, which provides ground-truth correspondences. 
Again, points are projected onto their nearest neighbor when discretizations are not compatible:
\begin{equation}
    \mathcal{E}_{corr}^{(s, t)}(p) = \frac{1}{\sqrt{S^t}}\text{dist}_{geo}(P(p), Q(p))
\end{equation}
where $p$ is a vertex on the mesh of the source object, $P(p)$ is the ground-truth correspondence of $p$ (projected onto the target), and $Q(p)$ is the predicted correspondence of $p$ on the target.
\cref{fig:quant:shape_matching} shows the cumulative error curve computed from $\mathcal{E}_{corr}^{(s, t)}$ on all intra-class pairs. 
In this experiment, our Laplacian does outperform the point cloud Laplacian again, but the extracted meshes from GOF perform slightly better. 
This could be due to smaller effects of artifacts in the mesh from the dimensionality reduction in functional maps.
Qualitative results can be found in \cref{fig:cmap:shape_matching}.

\subsection{Adaptive training} \label{sub:exp:adapted}
Remarkably, this training scheme leads to clean geometry with nearly no outlier splats in the geometry (see \cref{fig:filtering}) while maintaining the shape of the object. 
In addition to having the highest quality geometry in the sense of the spectrum of Laplacian operator, the adaptive training produces a significantly more compact representation (in terms of number of points) compared to the reconstructed meshes, the Gaussian splats from GOF, the post-processed 3D Gaussian splats, and even the original meshes, see \cref{table:size}. Moreover, the rendering quality is not notably degraded using adaptive training, see \cref{table:rendering}.

\section{Conclusion}
\label{sec:conclusion}

We have introduced a new Laplace-Beltrami operator that operates directly on 3D Gaussian splatting representations without conversion into intermediate formats. 
Our formulation is based on a Mahalanobis distance neighborhood computation which preserves the connectivity information implied by the anisotropic covariance matrices of each splat. 
Our method addresses a fundamental gap in geometry processing for Gaussian representations. 
Experimental validation demonstrates superior performance compared to point cloud approaches and competitive results with mesh-based methods across eigenvalue analysis, curvature estimation, geodesic computation, and shape correspondence. 
The direct computation eliminates mesh reconstruction overhead while our adaptive training produces remarkably compact and clean representations.
We hope this will lead to a broader adaption of 3DGS beyond rendering in computer vision and graphics applications. 

\paragraph{Limitations.} The quality of our LBO depends on the assumption that the splats are mostly flat and aligned to the surface. This is only the case for some 3DGS implementations but due to the wider adaption of geometry priors not an unreasonable assumption.

\vspace{5pt}
{\bf\noindent Acknowledgements.} This research has been funded by the DFG Sachbeihilfe grant \text{LA 5191/2-1}.
{
    \small
    \bibliographystyle{ieeenat_fullname}
    \bibliography{main}
}

\end{document}